\documentclass[sn-mathphys,iicol,Numbered]{sn-jnl}
\usepackage{graphicx}%
\usepackage{multirow}%
\usepackage{amsmath,amssymb,amsfonts}%
\usepackage{amsthm}%
\usepackage{mathrsfs}%
\usepackage[title]{appendix}%
\usepackage{xcolor}%
\usepackage{textcomp}%
\usepackage{manyfoot}%
\usepackage{booktabs}%
\usepackage{algorithm}%
\usepackage{algorithmicx}%
\usepackage{algpseudocode}%
\usepackage{listings}%
\usepackage{subcaption}
\usepackage[numbers]{natbib}
\usepackage{multibib}
\usepackage[switch]{lineno}
\newcites{app}{Appendix References}

\theoremstyle{thmstyleone}%
\theoremstyle{thmstyletwo}%

\theoremstyle{thmstylethree}%
\usepackage[normalem]{ulem}  

\graphicspath{{figs/}{./}}
\newcommand{\vect}[1]{\boldsymbol{#1}}
\newcommand{\avg}[1]{\left\langle #1 \right\rangle}

\usepackage{tikz,xcolor,hyperref}
\definecolor{lime}{HTML}{A6CE39}
\DeclareRobustCommand{\orcidicon}{
	\begin{tikzpicture}
	\draw[lime, fill=lime] (0,0) 
	circle [radius=0.16] 
	node[white] {{\fontfamily{qag}\selectfont \tiny ID}};
	\draw[white, fill=white] (-0.0625,0.095) 
	circle [radius=0.007];
	\end{tikzpicture}
	\hspace{-2mm}
}
\foreach \x in {A, ..., Z}{%
	\expandafter\xdef\csname orcid\x\endcsname{\noexpand\href{https://orcid.org/\csname orcidauthor\x\endcsname}{\noexpand\orcidicon}}
}

\begin{document}
\title[Article Title]{
A Barnett-Like  Effect in Relativistic Heavy-Ion Collisions
}

\author[1,2,3]{\fnm{Dai-Neng} \sur{Liu}\orcidA{}}        
\author*[1,2]{\fnm{Kai-Jia}   \sur{Sun}\orcidC{}}        \email{kjsun@fudan.edu.cn}
\author*[5,6]{\fnm{Jan}       \sur{Steinheimer}\orcidB{}}\email{j.steinheimer-froschauer@gsi.de}
\author*[1,2]{\fnm{Jin-Hui}   \sur{Chen}\orcidD{}}       \email{chenjinhui@fudan.edu.cn}
\author*[1,2,7]{\fnm{Yu-Gang} \sur{Ma}\orcidE{}}         \email{mayugang@fudan.edu.cn}
\author*[3,4]{\fnm{Marcus}    \sur{Bleicher}\orcidF{}}   \email{bleicher@itp.uni-frankfurt.de}

\affil[1]{\orgdiv{Key Laboratory of Nuclear Physics and Ion-beam Application (MOE), Institute of Modern Physics}, \orgname{Fudan University}, \orgaddress{\city{Shanghai}, \postcode{200433}, \country{China}}}

\affil[2]{\orgdiv{Shanghai Research Center for Theoretical Nuclear Physics}, \orgname{NSFC and Fudan University}, \orgaddress{\city{Shanghai}, \postcode{200438}, \country{China}}}

\affil[3]{\orgdiv{Institut f\"ur Theoretische Physik}, \orgname{Goethe-Universit\"at Frankfurt}, \orgaddress{\street{Max-von-Laue-Str.~1}, \postcode{D-60438}, \city{Frankfurt am Main}, \country{Germany}}}

\affil[4]{\orgdiv{Helmholtz Research Academy Hesse for FAIR (HFHF)}, \orgname{GSI Helmholtzzentrum f\"ur Schwerionenforschung GmbH}, \orgaddress{\street{Campus Frankfurt, Max-von-Laue-Str.~12}, \postcode{60438}, \city{Frankfurt am Main}, \country{Germany}}}

\affil[5]{\orgname{GSI Helmholtzzentrum f\"ur Schwerionenforschung GmbH}, \orgaddress{\street{Planckstr.~1}, \postcode{D-64291}, \city{Darmstadt}, \country{Germany}}}

\affil[6]{\orgname{Frankfurt Institute for Advanced Studies}, \orgaddress{\street{Ruth-Moufang-Str.~1}, \postcode{60438}, \city{Frankfurt am Main}, \country{Germany}}}

\affil[7]{\orgdiv{School of Physics}, \orgname{East China Normal University}, \orgaddress{\city{Shanghai}, \postcode{200241}, \country{China}}}

\date{\today} 
\abstract{ The Barnett effect refers to the spontaneous magnetization or polarization of rotating matter via spin–orbit coupling, a fundamental phenomenon observed across a multitude of physical systems down to atomic scales. Here, we demonstrate that this effect extends to subatomic regimes within relativistic heavy-ion collisions, where the most vortical fluid ever created in terrestrial laboratories is produced and the orbital angular momentum is transferred to emitted hadrons via spin-vorticity coupling. By performing transport simulations  spanning a broad energy range, we discover a  nearly-linear scaling relation between the global spin polarization of $\Lambda$ hyperons   and the source tilt angle extracted via two-pion femtoscopic correlations.   This scaling reveals that global $\Lambda$ polarization induced by the local vorticity field is modulated by the inclined geometry of the produced strongly-interacting nuclear matter   in these collisions. Supported by existing experimental measurements, our results unravel a geometry-modulated Barnett-like effect at the femtometre ($10^{-15}$ m) scale.  
 }



\maketitle
The unification of special relativity and quantum mechanics gives birth to modern quantum field theory that   predicts the coupling of a particle's orbital motion with its intrinsic spin~\cite{Schwartz2014}. The macroscopic manifestation of this spin-orbit coupling is the Barnett effect~\cite{Barnett,doi:10.1126/science.48.1239.303} in which global mechanical rotation magnetizes a rigid body.
The Barnett effect has been observed in a wide variety of macroscopic (i.e. atomic scale and larger) physical systems, extending from fast rotating water~\cite{PhysRevLett.122.177202} and optically driven phonons~\cite{Davies2024Phononic} to quantum spin  gyroscopes~\cite{Wood_2017} and   rotators~\cite{zjpt-whpf}.

\begin{figure}
    \centering
    \includegraphics[width=0.5\textwidth]{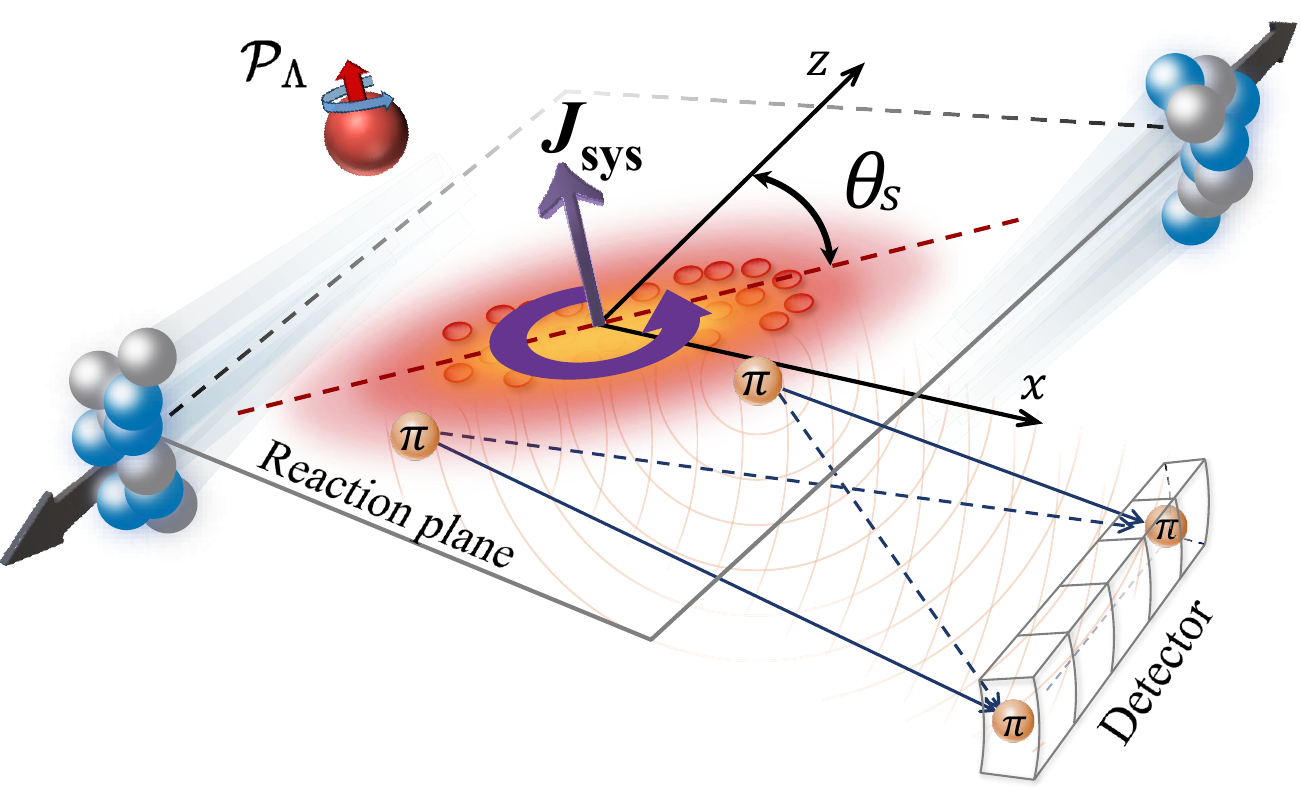} 
    \caption{\textbf{Schematic illustration of semi-central relativistic heavy-ion collisions.}  
    Two spectator fragments propagate along the beam ($z$) axis, leaving strongly-interacting vortical matter near mid-rapidity. The average vorticity points along the total orbital angular momentum $J_{\rm sys}$, perpendicular to the reaction ($z$--$x$) plane.  Spinning tops represent $\Lambda$ hyperons, and $\theta_s$ denotes the tilt of the emission source relative to the beam direction. The purple curled arrow indicates the direction of $\mathbf{J}_{\mathrm{sys}}$, rather than a global rotational flow. Experimentally, the reaction plane is reconstructed from the deflection of fragments in beam--beam counters. The global $\Lambda$ polarization ($\mathcal{P}_\Lambda$) is determined from the angular distribution of  $\Lambda$'s parity-violating weak decay, $\Lambda\rightarrow\pi^-+p$, whereas  $\theta_s$ is  extracted from momentum correlations arising from quantum interference between identical pions measured by the detector.
    }
    \label{fig:schematic}
\end{figure}

An analogous phenomenon occurs in high-energy nuclear reactions, where heavy atomic nuclei collide at nearly the speed of light. Figure \ref{fig:schematic} schematically illustrates a semi-central relativistic heavy-ion collision, with the beam axis along $z$, the impact parameter along $x$, and the $y$ direction being orthogonal to the reaction plane spanned by $x$ and $z$. The collision creates strongly interacting nuclear matter at femtometre scales and temperatures exceeding $T\sim10^{12}$~K~\cite{Andronic:2017pug}, with a source tilted by an angle $\theta_s$ relative to the beam direction and carrying substantial orbital angular momentum, $J_\text{sys}\sim10^3\hbar$ with $\hbar$ denoting the reduced Planck constant, perpendicular to the reaction plane.  
Through the spin–orbit or spin–vorticity coupling,  part of this angular momentum is transferred to the spin degrees of freedom of hadrons, giving rise to the phenomenon of global spin polarization~\cite{Liang:2004ph,Voloshin:2004ha}. Measurements of $\Lambda$ and $\bar{\Lambda}$ hyperon polarizations across various collision energies~\cite{STAR:2017ckg,ALICE:2019onw,HADES:2022enx}, supported by hydrodynamic~\cite{Karpenko:2016jyx,Xie:2017upb,Fu:2020oxj,Ivanov:2025izv} and transport~\cite{Jiang:2016woz,Li:2017slc,Shi:2017wpk,Deng:2021miw} model calculations, have demonstrated that the produced strongly-interacting matter is the most vortical fluid ever produced in terrestrial laboratories~\cite{Becattini:2020ngo,Niida:2024ntm,Becattini:2024uha,Chen:2024afy}. 

Unlike rigid rotators, however, the expanding fireball in relativistic heavy-ion collisions develops a non-uniform vorticity field, rendering global spin polarization sensitive to both local  vorticity  and overall source geometry. Directly probing these features remains challenging because neither the local vorticity field nor the particle emission coordinates are directly observable.

A promising way to overcome this limitation is through two-particle femtoscopic correlations, known as femtoscopy or Hanbury-Brown--Twiss (HBT) interferometry~\cite{HanburyBrown:1956bqd,Pratt:1984su}, which exploit quantum interference between identical hadrons to probe the size and orientation of the emission source in heavy-ion collisions~\cite{Heinz:1999rw,Lisa:2005dd}. In particular, azimuthally sensitive correlations of identical pion pairs can determine the tilt angle of the emission ellipsoid in the reaction plane~\cite{Lisa:2000ip,Mount:2010ey,Lisa:2011na,Khyzhniak:2024chj}.

\begin{figure*}[!t]
    \centering
\includegraphics[width=0.46\textwidth]{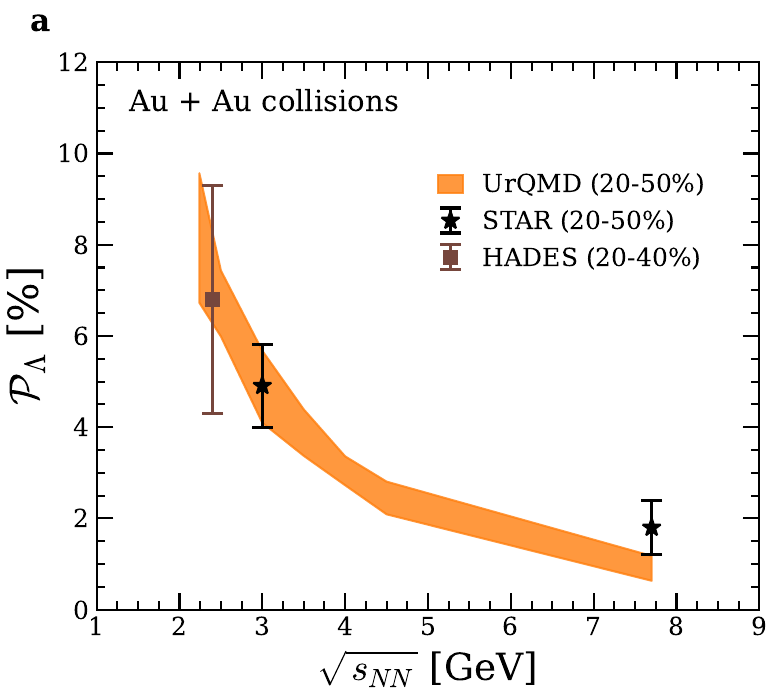} 
\includegraphics[width=0.46\textwidth]{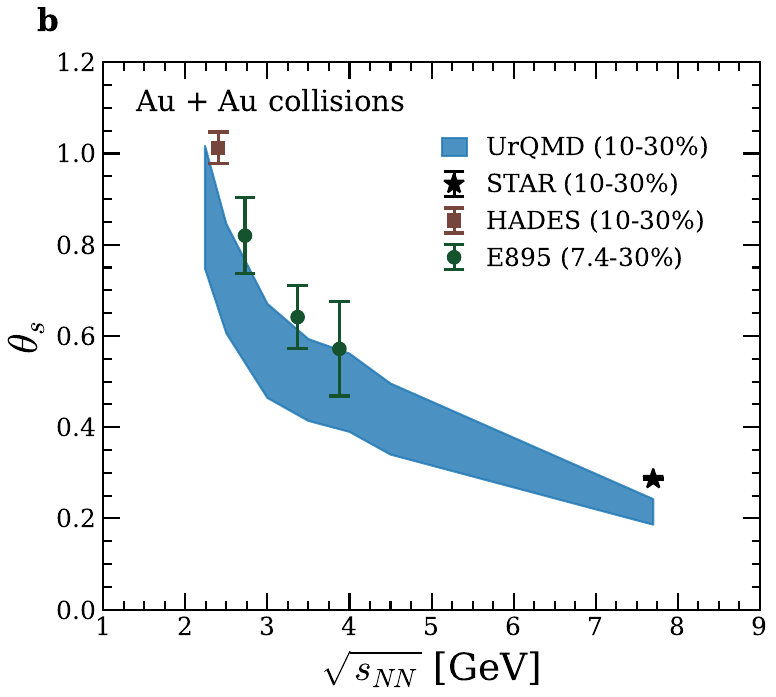} %
    \caption{\textbf{Collision-energy dependence of (a) global $\Lambda$ polarization ($\mathcal{P}_\Lambda$) and (b) source tilt angle ($\theta_s$ in radian) in Au+Au collisions.} Theoretical results for global $\Lambda$ polarization (orange band) are obtained with UrQMD for 20--50\% Au+Au collisions in mid-rapidity $|y|<1$ at $\sqrt{s_{NN}}=2.24$--7.7 GeV . The source tilt angle (blue band) is calculated at 10--30\% centrality over the same energy range.  Experimental data, with combined statistical and systematic uncertainties, are from the E895 \cite{E895:2000opr}, HADES~\cite{HADES:2019lek,HADES:2022enx}, and STAR~\cite{STAR:2017ckg,STAR:2021beb,STAR:2026skb} Collaborations. For consistency with other measurements, $\pi/2$ is added to the HADES tilt angle in Eq.~(\ref{eq:hbttilt}) for both theory and data~\cite{STAR:2026skb}.}
    \label{fig:enerDep}
\end{figure*}

Here, we use the Ultra-relativistic Quantum Molecular Dynamics (UrQMD) model~\cite{Bass:1998ca,Bleicher:1999xi} to investigate the global $\Lambda$ polarization ($\mathcal{P}_\Lambda$) and source tilt angle ($\theta_s$) in Au+Au collisions at $\sqrt{s_{{NN}}}=2.24$--$7.7~\mathrm{GeV}$. We find a nearly linear scaling between $\mathcal{P}_\Lambda$ and $\theta_s$ across the entire energy range, which is well supported by recent experimental measurements. This direct connection between spin polarization and emission geometry reveals a geometry-modulated Barnett-like effect in which the global spin response governed jointly by local thermal   vorticity and source geometry, rather than by a  single rigid-body angular velocity.

To show this connection, we first study the spin polarization of $\Lambda$ hyperons, which is commonly determined by the thermal vorticity of the produced strongly-interacting matter under the assumption of local thermal equilibrium~\cite{Becattini:2013fla}. The spin polarization vector of a particle with   spin $s$, mass $m$, four-momentum $p^\mu$ ($\mu=t,x,y,z$), and emitted at space-time point $x^\mu$, is given by~\cite{Becattini:2013fla}
\begin{equation}
     \mathcal{P}^\mu(x,p)=-\frac{s+1}{6m}(1-n_F)\epsilon^{\mu\nu\rho\sigma}p_\nu\varpi_{\rho\sigma}(x),
     \label{eq:pola}
\end{equation}
where $n_F = 1/[1 + \exp((u \cdot p \mp \mu)/T)]$ is the Fermi-Dirac distribution function for particles $(-)$ and antiparticles $(+)$. In heavy-ion collisions, $1-n_F \approx 1$ at the kinetic freeze-out due to the low densities.  
In Eq.~\eqref{eq:pola}, the thermal vorticity tensor is defined as ${\varpi_{\mu\nu}=-\frac{1}{2}(\partial_\mu\beta_\nu-\partial_\nu\beta_\mu)}$ with $\beta^\mu=u^\mu/T$ being the four-velocity (satisfying the normalization condition $u^\mu u_\mu=1$) of the fluid divided by temperature. In the UrQMD simulation, the temperature field $T(x)$ and velocity field $u(x)$ are obtained by solving Landau-frame
equations with corresponding equation of state~\cite{Liu:2026lny}. Note that these fields are determined by averaging over a sufficiently large number (typically $10^5$) of events from UrQMD and the $\Lambda$ spin polarization  is calculated in its rest frame with a Lorentz boost~\cite{Becattini:2013fla} (see  \hyperref[methods:polarization]{Methods}  for details).

Figure~\ref{fig:enerDep}(a) compares the beam-energy dependence of the global $\Lambda$ polarization ($\mathcal{P}_\Lambda$) measured in experiments with our theoretical calculations. The orange bands show UrQMD results for 20--50\% Au+Au collisions at $\sqrt{s_{\rm NN}}=2.24$--$7.7~\mathrm{GeV}$ within mid-rapidity $|y|<1$. Solid symbols represent measurements from the HADES~\cite{HADES:2022enx} and STAR~\cite{STAR:2017ckg,STAR:2021beb} Collaborations. The polarization decreases monotonically over the energy range studied, and the UrQMD calculations agree well with the experimental data. Our results are also consistent with other model calculations summarized in Refs.~\cite{Becattini:2020ngo,HADES:2022enx,Niida:2024ntm}.

The particle emission coordinates and source geometry cannot be measured directly in experiments, where only the final-state particle four-momenta are accessible. Two-particle femtoscopic momentum correlations provide a means of reconstructing the space-time geometry of the emission source in heavy-ion collisions~\cite{Lisa:2000ip,Lisa:2005dd}. Using a correlation afterburner approach (CRAB)~\cite{Pratt:1994uf}, the two-particle momentum correlation function, defined as the ratio of the  two-particle inclusive and single-particle inclusive spectra, is calculated as~\cite{Lisa:2005dd}, 
\begin{equation}
C(\vect k,\vect K)=\frac{\int d\Sigma ~S_1(x_1,\vect p_1)S_2(x_2,\vect p_2)|\psi_{\vect k}(\vect x)|^2}{\int d\Sigma~ S_1(x_1,\vect p_1)S_2(x_2,\vect p_2)},
\label{eq:hbtcorr0}
\end{equation}
where $d\Sigma=d^4 x_1 d^4 x_2$ denotes two-particle volume element;  $S(x,\vect p)$ denotes the source function at given freeze-out space-time coordinate $x$ and three-momentum $\vect p$;  $\vect k=\vect p_1 - \vect p_2$ and   $\vect K = (\vect p_1 + \vect p_2)/2$  are the relative and average pair momenta, respectively, in the   ``out-side-long'' coordinate system, where the relative momentum is decomposed into components parallel to the beam ($l$ = longitudinal), parallel to the transverse momentum $\vect K_T$ ($o$ = out), and in the remaining third direction ($s$ = side)~\cite{Lisa:2000ip}; $\psi_{\vect k}(\vect x)$ denotes the outgoing wavefunction with $\vect x$ being the relative   position in the pair rest frame, obtained by propagating the earlier-emitted particle to the emission time of the later one. 

For identical pions emitted from an isotropic Gaussian source of radius $R$, quantum interference gives $|\psi_{\vect k}(\vect x)|^2=1+\cos(\vect k\cdot \vect x)$, leading to $C(\vect k)=1+\exp(-R^2k^2)$~\cite{Lisa:2005dd}. Thus, the momentum dependence of the correlation function directly encodes the spatial extent of the emitting source.

For an expanding anisotropic  fireball in realistic heavy-ion collisions, the  correlation function is often approximated as
\begin{equation}
C(\vect k,\vect K, \phi)=1+\lambda(\vect K)\exp\left[-\sum_{i,j=o,s,l}k_ik_jR_{ij}^2(\vect K, \phi)\right],
\label{eq:hbtcorr}
\end{equation}
where  $\phi$ represents the azimuthal angle between the direction of pair transverse momentum $\vect K_T$ and  the impact parameter, $\lambda$ denotes the correlation strength parameter, and $R^2_{ij}$ are the HBT radii matrix. The tilt angle of the homogeneity region can then be extracted from the azimuthal dependence of these radii~\cite{Lisa:2000ip,Lisa:2011na},
\begin{equation}
\theta_{s}\approx\frac{1}{2}\tan^{-1}\!\left(
\frac{-4 R_{sl,1}^2}{R_{l,0}^2-R_{s,0}^2+2  R_{s,2}^2}
\right),
\label{eq:hbttilt}
\end{equation}
where the subscripts denote the corresponding Fourier harmonics (see  \hyperref[methods:tilt]{Methods} for details).

Figure~\ref{fig:enerDep}(b) shows the source tilt angle ($\theta_s$) calculated within UrQMD for semi-central (10--30\%) Au+Au collisions at  mid-rapidity $|y|<1$ (blue band). The tilt angle is obtained via Eq.~(\ref{eq:hbttilt}) with the HBT radii obtained by fitting the two-pion correlation function in Eq.~(\ref{eq:hbtcorr}). It decreases monotonically with increasing collision energy. At lower beam energies, strong baryon stopping produces substantial interpenetration shear between projectile and target matter, yielding a highly inclined emission ellipsoid. As the beam energy increases,   longitudinal Bjorken expansion becomes dominant, stretching the source along the beam direction and reducing the spatial tilt. This trend agrees closely with experimental tilt angles extracted from azimuthally sensitive two-pion femtoscopy by E895 at AGS~\cite{E895:2000opr}, HADES~\cite{HADES:2019lek}, and STAR~\cite{STAR:2026skb}.  The centrality selections for polarization (20–50\%) 
and tilt (10–30\%) follow those used in the experimental comparisons. These modest differences   have only a minor effect on the polarization–tilt   correlation and do not alter our conclusion. For comparison with experimental data, the theoretical tilt angle evaluated at low  pair transverse-momentum interval of $0.15<K_T<0.20$ GeV/$c$ at $\sqrt{s_{NN}}> 3$ GeV and  $0.05<K_T<0.10$ GeV/$c$ at $\sqrt{s_{NN}} \le 3$ GeV.

\begin{figure}[!t]
    \centering
    \includegraphics[width=1\columnwidth]{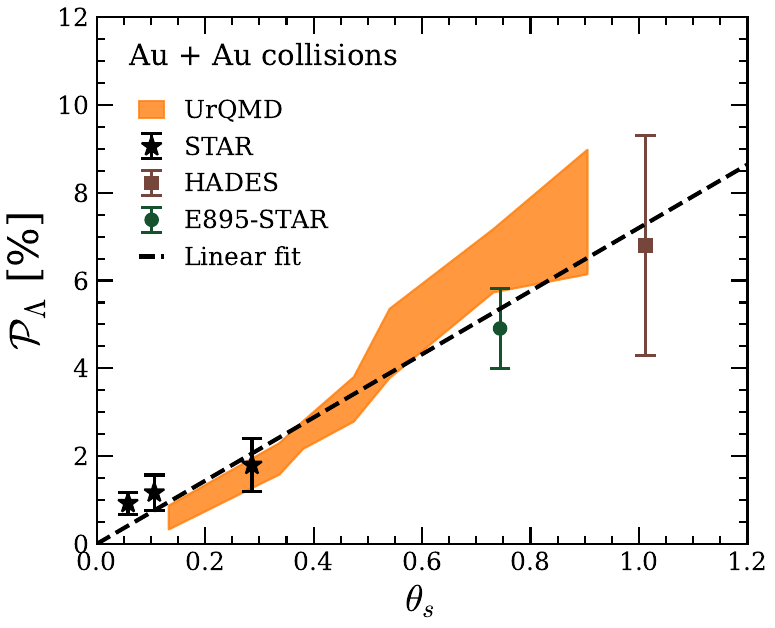} 
    \caption{\textbf{Polarization-tilt correlation in relativistic heavy-ion collisions.}  Global $\Lambda$ polarization ($\mathcal{P}_\Lambda$) versus the source tilt angle ($\theta_s$). Theoretical results shown by the orange band are obtained with UrQMD for  Au+Au collisions at $\sqrt{s_{NN}}=2.24$--7.7 GeV with 20--50\% centrality for $\mathcal{P}_\Lambda$ and 10--30\% for $\theta_s$. Experimental data of $\mathcal{P}_\Lambda$, with combined statistical and systematic uncertainties, are taken from Refs.~\cite{E895:2000opr,STAR:2017ckg,HADES:2019lek,STAR:2021beb,HADES:2022enx,STAR:2026skb}. For the data point at $\sqrt{s_{NN}}=3$ GeV denoted by E895-STAR, the source tilt angle is derived by interpolating the values at the neighboring AGS energies, 2.73 and 3.37 GeV, while the polarization is taken from STAR measurements. The black solid stars denote STAR data, including additional two preliminary points at $\sqrt{s_{NN}}=14.5$ and 27 GeV~\cite{STAR:2017ckg,STAR:2026skb}. A linear fit to the experimental data is shown by the black dashed line. }
    \label{fig:barnett}
\end{figure}

Figure~\ref{fig:barnett} further displays   the global $\Lambda$ polarization against the source tilt angle. Theoretical results for Au+Au collisions at $\sqrt{s_{NN}}=2.24$--7.7 GeV from UrQMD simulations (orange bands)  exhibit a nearly linear scaling, consistent within uncertainties with measurements from E895~\cite{E895:2000opr}, HADES~\cite{HADES:2019lek,HADES:2022enx}, and STAR~\cite{STAR:2017ckg,STAR:2021beb,STAR:2026skb}. The two additional STAR data points (solid black stars) at $\sqrt{s_{\mathrm{NN}}}=14.5$ and $27~\mathrm{GeV}$~\cite{STAR:2017ckg,STAR:2026skb}, beyond the energy range of the present transport calculation, also follow this trend. For the E895-STAR data point at $\sqrt{s_{NN}}=3$ GeV, the source tilt angle is obtained by interpolating between the neighboring AGS energies of  $\sqrt{s_{NN}}=$2.73 and 3.37 GeV, while the polarization is taken from STAR measurements. A linear fit to the experimental data yields an even tighter linear relation (black dashed line), i.e.,
\begin{equation}
    \mathcal{P}_\Lambda = (0.072 \pm 0.010) \, \theta_s, \label{eq:linear}
\end{equation}
where $\theta_s$ is expressed in radians and $\mathcal{P}_\Lambda$ as a 
fraction rather than a percentage. The quoted  fit includes the polarization uncertainties but  neglects uncertainties in $\theta_s$.

The underlying mechanism of the relation in Eq. (\ref{eq:linear}) can be traced to the quadrupolar structure of the vorticity, whose sign alternates across the four quadrants, as shown in  \hyperref[methods:linear]{Methods}. In the first and third quadrants ($xz>0$), the vorticity aligns with the global angular momentum, whereas in the other two quadrants it points in the opposite direction. Their contributions to the global $\Lambda$ polarization in Eq.~(\ref{eq:pola}) therefore largely cancel, with the cancellation becoming increasingly complete at higher collision energies as the system approaches longitudinal boost invariance~\cite{Li:2017slc,Shi:2017wpk}. A finite source tilt breaks this cancellation, leaving a residual polarization that is proportional to $\theta_s$ at leading order, i.e. $\mathcal{P}_\Lambda\propto |\varpi_{zx}|\theta_s$ where $|\varpi_{zx}|$ denotes the magnitude of local thermal vorticity (see \hyperref[methods:linear]{Methods}). This geometric scaling distinguishes the response from the conventional Barnett effect in rigidly rotating matter, where the polarization is primarily governed by the angular velocity divided by temperature. It also naturally explains the decrease of $\mathcal{P}_\Lambda$ with increasing collision energy as the source tilt vanishes toward longitudinal boost invariance.  The agreement between theoretical prediction by UrQMD and experimental data thus provides evidence for a geometry-modulated Barnett-like effect occurring  at  subatomic scales down to femtometre ($10^{-15}$~m) and extreme  temperatures exceeding $10^{12}$ K.

Joint precision measurements of $\Lambda$ polarization and the source tilt angle can further test this geometry-modulated spin response. In particular, measurements using unified centrality selections and specified momentum acceptances will sharpen quantitative comparisons with theoretical models. Extending our framework across a broader range of collision energies, from a few GeV to the TeV scale, will deepen our understanding of the interplay among spin polarization, vorticity, and source geometry. More broadly, applying femtoscopic techniques to other spin observables may help clarify the microscopic origin of spin–spin correlations~\cite{STAR:2025njp,STAR:2022fan,Zhang:2026tqs} and probe the chiral phase transition~\cite{Kiefer:2025xdp} in high-energy nuclear collisions.

\bmhead{Acknowledgments}
This work was supported in part by the National Key Research and Development Project of China under Grant No.~2024YFA1612500, the National Natural Science Foundation of China under Grants No.~12422509, No.~12375121, No.~12625501, No.~124B2102, 12688101 and No.~12547102, and the STCSM under Grant No. 23590780100 and 23JC1400200. 
D.-N.~Liu is also supported by the China Scholarship Council under Grant No.~202506100099. The computations in this research were performed using the CFFF platform of Fudan University.  

\bmhead{Data availability}
All the data supporting the findings in this work are available within the manuscript and any additional data are available from the corresponding authors upon reasonable request.

\bmhead{Code availability}
Inquiries about the code in this work will be responded to by the corresponding authors.

\bmhead{Author contributions}
D. N.~Liu and K. J. Sun performed the numerical simulations and prepared the figures.   All authors participated in discussions and contributed to the preparation of the manuscript.

\bmhead{Competing interests}
The authors declare no competing interests.


\begin{thebibliography}{51}
\ifx \bisbn   \undefined \def \bisbn  #1{ISBN #1}\fi
\ifx \binits  \undefined \def \binits#1{#1}\fi
\ifx \bauthor  \undefined \def \bauthor#1{#1}\fi
\ifx \batitle  \undefined \def \batitle#1{#1}\fi
\ifx \bjtitle  \undefined \def \bjtitle#1{#1}\fi
\ifx \bvolume  \undefined \def \bvolume#1{\textbf{#1}}\fi
\ifx \byear  \undefined \def \byear#1{#1}\fi
\ifx \bissue  \undefined \def \bissue#1{#1}\fi
\ifx \bfpage  \undefined \def \bfpage#1{#1}\fi
\ifx \blpage  \undefined \def \blpage #1{#1}\fi
\ifx \burl  \undefined \def \burl#1{\textsf{#1}}\fi
\ifx \doiurl  \undefined \def \doiurl#1{\url{https://doi.org/#1}}\fi
\ifx \betal  \undefined \def \betal{\textit{et al.}}\fi
\ifx \binstitute  \undefined \def \binstitute#1{#1}\fi
\ifx \binstitutionaled  \undefined \def \binstitutionaled#1{#1}\fi
\ifx \bctitle  \undefined \def \bctitle#1{#1}\fi
\ifx \beditor  \undefined \def \beditor#1{#1}\fi
\ifx \bpublisher  \undefined \def \bpublisher#1{#1}\fi
\ifx \bbtitle  \undefined \def \bbtitle#1{#1}\fi
\ifx \bedition  \undefined \def \bedition#1{#1}\fi
\ifx \bseriesno  \undefined \def \bseriesno#1{#1}\fi
\ifx \blocation  \undefined \def \blocation#1{#1}\fi
\ifx \bsertitle  \undefined \def \bsertitle#1{#1}\fi
\ifx \bsnm \undefined \def \bsnm#1{#1}\fi
\ifx \bsuffix \undefined \def \bsuffix#1{#1}\fi
\ifx \bparticle \undefined \def \bparticle#1{#1}\fi
\ifx \barticle \undefined \def \barticle#1{#1}\fi
\bibcommenthead
\ifx \bconfdate \undefined \def \bconfdate #1{#1}\fi
\ifx \botherref \undefined \def \botherref #1{#1}\fi
\ifx \url \undefined \def \url#1{\textsf{#1}}\fi
\ifx \bchapter \undefined \def \bchapter#1{#1}\fi
\ifx \bbook \undefined \def \bbook#1{#1}\fi
\ifx \bcomment \undefined \def \bcomment#1{#1}\fi
\ifx \oauthor \undefined \def \oauthor#1{#1}\fi
\ifx \citeauthoryear \undefined \def \citeauthoryear#1{#1}\fi
\ifx \endbibitem  \undefined \def \endbibitem {}\fi
\ifx \bconflocation  \undefined \def \bconflocation#1{#1}\fi
\ifx \arxivurl  \undefined \def \arxivurl#1{\textsf{#1}}\fi
\csname PreBibitemsHook\endcsname

\bibitem[\protect\citeauthoryear{Schwartz}{2014}]{Schwartz2014}
\begin{bbook}
\bauthor{\bsnm{Schwartz}, \binits{M.D.}}:
\bbtitle{Quantum Field Theory and the Standard Model},
\bedition{1st} edn.,
p. \bfpage{850}.
\bpublisher{Cambridge University Press},
\blocation{Cambridge}
(\byear{2014})
\end{bbook}
\endbibitem

\bibitem[\protect\citeauthoryear{Barnett}{1915}]{Barnett}
\begin{barticle}
\bauthor{\bsnm{Barnett}, \binits{S.J.}}:
\batitle{Magnetization by rotation}.
\bjtitle{Phys. Rev.}
\bvolume{6},
\bfpage{239}--\blpage{270}
(\byear{1915})
\doiurl{10.1103/PhysRev.6.239}
\end{barticle}
\endbibitem

\bibitem[\protect\citeauthoryear{Barnett}{1918}]{doi:10.1126/science.48.1239.303}
\begin{barticle}
\bauthor{\bsnm{Barnett}, \binits{S.J.}}:
\batitle{Magnetization by rotation}.
\bjtitle{Science}
\bvolume{48}(\bissue{1239}),
\bfpage{303}--\blpage{309}
(\byear{1918})
\doiurl{10.1126/science.48.1239.303}
\end{barticle}
\endbibitem

\bibitem[\protect\citeauthoryear{Arabgol and
  Sleator}{2019}]{PhysRevLett.122.177202}
\begin{barticle}
\bauthor{\bsnm{Arabgol}, \binits{M.}},
\bauthor{\bsnm{Sleator}, \binits{T.}}:
\batitle{Observation of the nuclear barnett effect}.
\bjtitle{Phys. Rev. Lett.}
\bvolume{122},
\bfpage{177202}
(\byear{2019})
\doiurl{10.1103/PhysRevLett.122.177202}
\end{barticle}
\endbibitem

\bibitem[\protect\citeauthoryear{Davies et~al.}{2024}]{Davies2024Phononic}
\begin{barticle}
\bauthor{\bsnm{Davies}, \binits{C.S.}},
\bauthor{\bsnm{Fennema}, \binits{F.G.N.}},
\bauthor{\bsnm{Tsukamoto}, \binits{A.}},
\bauthor{\bsnm{Razdolski}, \binits{I.}},
\bauthor{\bsnm{Kimel}, \binits{A.V.}},
\bauthor{\bsnm{Kirilyuk}, \binits{A.}}:
\batitle{Phononic switching of magnetization by the ultrafast {Barnett}
  effect}.
\bjtitle{Nature}
\bvolume{628}(\bissue{8008}),
\bfpage{540}--\blpage{544}
(\byear{2024})
\doiurl{10.1038/s41586-024-07200-x}
\end{barticle}
\endbibitem

\bibitem[\protect\citeauthoryear{Wood et~al.}{2017}]{Wood_2017}
\begin{barticle}
\bauthor{\bsnm{Wood}, \binits{A.A.}},
\bauthor{\bsnm{Lilette}, \binits{E.}},
\bauthor{\bsnm{Fein}, \binits{Y.Y.}},
\bauthor{\bsnm{Perunicic}, \binits{V.S.}},
\bauthor{\bsnm{Hollenberg}, \binits{L.C.L.}},
\bauthor{\bsnm{Scholten}, \binits{R.E.}},
\bauthor{\bsnm{Martin}, \binits{A.M.}}:
\batitle{Magnetic pseudo-fields in a rotating electron–nuclear spin system}.
\bjtitle{Nature Physics}
\bvolume{13}(\bissue{11}),
\bfpage{1070}--\blpage{1073}
(\byear{2017})
\doiurl{10.1038/nphys4221}
\end{barticle}
\endbibitem

\bibitem[\protect\citeauthoryear{Wachter et~al.}{2026}]{zjpt-whpf}
\begin{barticle}
\bauthor{\bsnm{Wachter}, \binits{V.}},
\bauthor{\bsnm{Kusminskiy}, \binits{S.V.}},
\bauthor{\bsnm{H\'etet}, \binits{G.}},
\bauthor{\bsnm{Stickler}, \binits{B.A.}}:
\batitle{Gyroscopically stabilized quantum spin rotors}.
\bjtitle{Phys. Rev. Lett.}
\bvolume{136},
\bfpage{073604}
(\byear{2026})
\doiurl{10.1103/zjpt-whpf}
\end{barticle}
\endbibitem

\bibitem[\protect\citeauthoryear{Andronic et~al.}{2018}]{Andronic:2017pug}
\begin{barticle}
\bauthor{\bsnm{Andronic}, \binits{A.}},
\bauthor{\bsnm{Braun-Munzinger}, \binits{P.}},
\bauthor{\bsnm{Redlich}, \binits{K.}},
\bauthor{\bsnm{Stachel}, \binits{J.}}:
\batitle{{Decoding the phase structure of QCD via particle production at high
  energy}}.
\bjtitle{Nature}
\bvolume{561}(\bissue{7723}),
\bfpage{321}--\blpage{330}
(\byear{2018})
\doiurl{10.1038/s41586-018-0491-6}
\end{barticle}
\endbibitem

\bibitem[\protect\citeauthoryear{Liang and Wang}{2005}]{Liang:2004ph}
\begin{barticle}
\bauthor{\bsnm{Liang}, \binits{Z.-T.}},
\bauthor{\bsnm{Wang}, \binits{X.-N.}}:
\batitle{{Globally polarized quark-gluon plasma in non-central A+A
  collisions}}.
\bjtitle{Phys. Rev. Lett.}
\bvolume{94},
\bfpage{102301}
(\byear{2005})
\doiurl{10.1103/PhysRevLett.94.102301} .
\bcomment{[Erratum: Phys.Rev.Lett. 96, 039901 (2006)]}
\end{barticle}
\endbibitem

\bibitem[\protect\citeauthoryear{Voloshin}{2004}]{Voloshin:2004ha}
\begin{botherref}
\oauthor{\bsnm{Voloshin}, \binits{S.A.}}:
{Polarized secondary particles in unpolarized high energy hadron-hadron
  collisions?}
(2004)
{\href{https://arxiv.org/abs/nucl-th/0410089}{{arXiv:nucl-th/0410089}}}
\end{botherref}
\endbibitem

\bibitem[\protect\citeauthoryear{Adam et~al.}{2017}]{STAR:2017ckg}
\begin{barticle}
\bauthor{\bsnm{Adam}, \binits{J.}}, \betal:
\batitle{{Global $\Lambda$ hyperon polarization in nuclear collisions: evidence
  for the most vortical fluid}}.
\bjtitle{Nature}
\bvolume{548},
\bfpage{62}--\blpage{65}
(\byear{2017})
\doiurl{10.1038/nature23004}
\end{barticle}
\endbibitem

\bibitem[\protect\citeauthoryear{Acharya et~al.}{2020}]{ALICE:2019onw}
\begin{barticle}
\bauthor{\bsnm{Acharya}, \binits{S.}}, \betal:
\batitle{{Global polarization of $\Lambda$ and $\bar \Lambda$ hyperons in Pb-Pb
  collisions at $\sqrt {s_{NN}}$ = 2.76 and 5.02 TeV}}.
\bjtitle{Phys. Rev. C}
\bvolume{101}(\bissue{4}),
\bfpage{044611}
(\byear{2020})
\doiurl{10.1103/PhysRevC.101.044611}
{\href{https://arxiv.org/abs/1909.01281}{{arXiv:1909.01281}}}
{[nucl-ex]}.
\bcomment{[Erratum: Phys.Rev.C 105, 029902 (2022)]}
\end{barticle}
\endbibitem

\bibitem[\protect\citeauthoryear{Abou~Yassine et~al.}{2022}]{HADES:2022enx}
\begin{barticle}
\bauthor{\bsnm{Abou~Yassine}, \binits{R.}}, \betal:
\batitle{{Measurement of global polarization of {\ensuremath{\Lambda}} hyperons
  in few-GeV heavy-ion collisions}}.
\bjtitle{Phys. Lett. B}
\bvolume{835},
\bfpage{137506}
(\byear{2022})
\doiurl{10.1016/j.physletb.2022.137506}
{\href{https://arxiv.org/abs/2207.05160}{{arXiv:2207.05160}}}
{[nucl-ex]}
\end{barticle}
\endbibitem

\bibitem[\protect\citeauthoryear{Karpenko and
  Becattini}{2017}]{Karpenko:2016jyx}
\begin{barticle}
\bauthor{\bsnm{Karpenko}, \binits{I.}},
\bauthor{\bsnm{Becattini}, \binits{F.}}:
\batitle{{Study of $\Lambda $ polarization in relativistic nuclear collisions
  at $\sqrt{s_\mathrm {NN}}=7.7$ \textendash{}200 GeV}}.
\bjtitle{Eur. Phys. J. C}
\bvolume{77}(\bissue{4}),
\bfpage{213}
(\byear{2017})
\doiurl{10.1140/epjc/s10052-017-4765-1}
\end{barticle}
\endbibitem

\bibitem[\protect\citeauthoryear{Xie et~al.}{2017}]{Xie:2017upb}
\begin{barticle}
\bauthor{\bsnm{Xie}, \binits{Y.}},
\bauthor{\bsnm{Wang}, \binits{D.}},
\bauthor{\bsnm{Csernai}, \binits{L.P.}}:
\batitle{{Global {\ensuremath{\Lambda}} polarization in high energy
  collisions}}.
\bjtitle{Phys. Rev. C}
\bvolume{95}(\bissue{3}),
\bfpage{031901}
(\byear{2017})
\doiurl{10.1103/PhysRevC.95.031901}
\end{barticle}
\endbibitem

\bibitem[\protect\citeauthoryear{Fu et~al.}{2021}]{Fu:2020oxj}
\begin{barticle}
\bauthor{\bsnm{Fu}, \binits{B.}},
\bauthor{\bsnm{Xu}, \binits{K.}},
\bauthor{\bsnm{Huang}, \binits{X.-G.}},
\bauthor{\bsnm{Song}, \binits{H.}}:
\batitle{{Hydrodynamic study of hyperon spin polarization in relativistic heavy
  ion collisions}}.
\bjtitle{Phys. Rev. C}
\bvolume{103}(\bissue{2}),
\bfpage{024903}
(\byear{2021})
\doiurl{10.1103/PhysRevC.103.024903}
\end{barticle}
\endbibitem

\bibitem[\protect\citeauthoryear{Ivanov}{2025}]{Ivanov:2025izv}
\begin{barticle}
\bauthor{\bsnm{Ivanov}, \binits{Y.B.}}:
\batitle{{Global {\ensuremath{\Lambda}} polarization in heavy-ion collisions at
  high baryon density}}.
\bjtitle{Phys. Rev. C}
\bvolume{112}(\bissue{1}),
\bfpage{014902}
(\byear{2025})
\doiurl{10.1103/qn56-fvnq}
{\href{https://arxiv.org/abs/2504.12200}{{arXiv:2504.12200}}}
{[nucl-th]}
\end{barticle}
\endbibitem

\bibitem[\protect\citeauthoryear{Jiang et~al.}{2016}]{Jiang:2016woz}
\begin{barticle}
\bauthor{\bsnm{Jiang}, \binits{Y.}},
\bauthor{\bsnm{Lin}, \binits{Z.-W.}},
\bauthor{\bsnm{Liao}, \binits{J.}}:
\batitle{{Rotating quark-gluon plasma in relativistic heavy ion collisions}}.
\bjtitle{Phys. Rev. C}
\bvolume{94}(\bissue{4}),
\bfpage{044910}
(\byear{2016})
\doiurl{10.1103/PhysRevC.94.044910}
{\href{https://arxiv.org/abs/1602.06580}{{arXiv:1602.06580}}}
{[hep-ph]}.
\bcomment{[Erratum: Phys.Rev.C 95, 049904 (2017)]}
\end{barticle}
\endbibitem

\bibitem[\protect\citeauthoryear{Li et~al.}{2017}]{Li:2017slc}
\begin{barticle}
\bauthor{\bsnm{Li}, \binits{H.}},
\bauthor{\bsnm{Pang}, \binits{L.-G.}},
\bauthor{\bsnm{Wang}, \binits{Q.}},
\bauthor{\bsnm{Xia}, \binits{X.-L.}}:
\batitle{{Global $\Lambda$ polarization in heavy-ion collisions from a
  transport model}}.
\bjtitle{Phys. Rev. C}
\bvolume{96}(\bissue{5}),
\bfpage{054908}
(\byear{2017})
\doiurl{10.1103/PhysRevC.96.054908}
\end{barticle}
\endbibitem

\bibitem[\protect\citeauthoryear{Shi et~al.}{2019}]{Shi:2017wpk}
\begin{barticle}
\bauthor{\bsnm{Shi}, \binits{S.}},
\bauthor{\bsnm{Li}, \binits{K.}},
\bauthor{\bsnm{Liao}, \binits{J.}}:
\batitle{{Searching for the Subatomic Swirls in the CuCu and CuAu Collisions}}.
\bjtitle{Phys. Lett. B}
\bvolume{788},
\bfpage{409}--\blpage{413}
(\byear{2019})
\doiurl{10.1016/j.physletb.2018.09.066}
{\href{https://arxiv.org/abs/1712.00878}{{arXiv:1712.00878}}}
{[nucl-th]}
\end{barticle}
\endbibitem

\bibitem[\protect\citeauthoryear{Deng et~al.}{2022}]{Deng:2021miw}
\begin{barticle}
\bauthor{\bsnm{Deng}, \binits{X.-G.}},
\bauthor{\bsnm{Huang}, \binits{X.-G.}},
\bauthor{\bsnm{Ma}, \binits{Y.-G.}}:
\batitle{{Lambda polarization in 108Ag+108Ag and 197Au+197Au collisions around
  a few GeV}}.
\bjtitle{Phys. Lett. B}
\bvolume{835},
\bfpage{137560}
(\byear{2022})
\doiurl{10.1016/j.physletb.2022.137560}
\end{barticle}
\endbibitem

\bibitem[\protect\citeauthoryear{Becattini and Lisa}{2020}]{Becattini:2020ngo}
\begin{barticle}
\bauthor{\bsnm{Becattini}, \binits{F.}},
\bauthor{\bsnm{Lisa}, \binits{M.A.}}:
\batitle{{Polarization and Vorticity in the Quark\textendash{}Gluon Plasma}}.
\bjtitle{Ann. Rev. Nucl. Part. Sci.}
\bvolume{70},
\bfpage{395}--\blpage{423}
(\byear{2020})
\doiurl{10.1146/annurev-nucl-021920-095245}
\end{barticle}
\endbibitem

\bibitem[\protect\citeauthoryear{Niida and Voloshin}{2024}]{Niida:2024ntm}
\begin{barticle}
\bauthor{\bsnm{Niida}, \binits{T.}},
\bauthor{\bsnm{Voloshin}, \binits{S.A.}}:
\batitle{{Polarization phenomenon in heavy-ion collisions}}.
\bjtitle{Int. J. Mod. Phys. E}
\bvolume{33}(\bissue{09}),
\bfpage{2430010}
(\byear{2024})
\doiurl{10.1142/S0218301324300108}
{\href{https://arxiv.org/abs/2404.11042}{{arXiv:2404.11042}}}
{[nucl-ex]}
\end{barticle}
\endbibitem

\bibitem[\protect\citeauthoryear{Becattini et~al.}{2024}]{Becattini:2024uha}
\begin{barticle}
\bauthor{\bsnm{Becattini}, \binits{F.}},
\bauthor{\bsnm{Buzzegoli}, \binits{M.}},
\bauthor{\bsnm{Niida}, \binits{T.}},
\bauthor{\bsnm{Pu}, \binits{S.}},
\bauthor{\bsnm{Tang}, \binits{A.-H.}},
\bauthor{\bsnm{Wang}, \binits{Q.}}:
\batitle{{Spin polarization in relativistic heavy-ion collisions}}.
\bjtitle{Int. J. Mod. Phys. E}
\bvolume{33}(\bissue{06}),
\bfpage{2430006}
(\byear{2024})
\doiurl{10.1142/9789811294679_0005}
\end{barticle}
\endbibitem

\bibitem[\protect\citeauthoryear{Chen et~al.}{2025}]{Chen:2024afy}
\begin{barticle}
\bauthor{\bsnm{Chen}, \binits{J.-H.}},
\bauthor{\bsnm{Liang}, \binits{Z.-T.}},
\bauthor{\bsnm{Ma}, \binits{Y.-G.}},
\bauthor{\bsnm{Sheng}, \binits{X.-L.}},
\bauthor{\bsnm{Wang}, \binits{Q.}}:
\batitle{{Vector meson{\textquoteright}s spin alignments in high energy
  reactions}}.
\bjtitle{Sci. China Phys. Mech. Astron.}
\bvolume{68}(\bissue{1}),
\bfpage{211001}
(\byear{2025})
\doiurl{10.1007/s11433-024-2495-1}
{\href{https://arxiv.org/abs/2407.06480}{{arXiv:2407.06480}}}
{[hep-ph]}
\end{barticle}
\endbibitem

\bibitem[\protect\citeauthoryear{Hanbury~Brown and
  Twiss}{1956}]{HanburyBrown:1956bqd}
\begin{barticle}
\bauthor{\bsnm{Hanbury~Brown}, \binits{R.}},
\bauthor{\bsnm{Twiss}, \binits{R.Q.}}:
\batitle{{A Test of a new type of stellar interferometer on Sirius}}.
\bjtitle{Nature}
\bvolume{178},
\bfpage{1046}--\blpage{1048}
(\byear{1956})
\doiurl{10.1038/1781046a0}
\end{barticle}
\endbibitem

\bibitem[\protect\citeauthoryear{Pratt}{1984}]{Pratt:1984su}
\begin{barticle}
\bauthor{\bsnm{Pratt}, \binits{S.}}:
\batitle{{Pion Interferometry for Exploding Sources}}.
\bjtitle{Phys. Rev. Lett.}
\bvolume{53},
\bfpage{1219}--\blpage{1221}
(\byear{1984})
\doiurl{10.1103/PhysRevLett.53.1219}
\end{barticle}
\endbibitem

\bibitem[\protect\citeauthoryear{Heinz and Jacak}{1999}]{Heinz:1999rw}
\begin{barticle}
\bauthor{\bsnm{Heinz}, \binits{U.W.}},
\bauthor{\bsnm{Jacak}, \binits{B.V.}}:
\batitle{{Two particle correlations in relativistic heavy ion collisions}}.
\bjtitle{Ann. Rev. Nucl. Part. Sci.}
\bvolume{49},
\bfpage{529}--\blpage{579}
(\byear{1999})
\doiurl{10.1146/annurev.nucl.49.1.529}
{\href{https://arxiv.org/abs/nucl-th/9902020}{{arXiv:nucl-th/9902020}}}
\end{barticle}
\endbibitem

\bibitem[\protect\citeauthoryear{Lisa et~al.}{2005}]{Lisa:2005dd}
\begin{barticle}
\bauthor{\bsnm{Lisa}, \binits{M.A.}},
\bauthor{\bsnm{Pratt}, \binits{S.}},
\bauthor{\bsnm{Soltz}, \binits{R.}},
\bauthor{\bsnm{Wiedemann}, \binits{U.}}:
\batitle{{Femtoscopy in relativistic heavy ion collisions}}.
\bjtitle{Ann. Rev. Nucl. Part. Sci.}
\bvolume{55},
\bfpage{357}--\blpage{402}
(\byear{2005})
\doiurl{10.1146/annurev.nucl.55.090704.151533}
{\href{https://arxiv.org/abs/nucl-ex/0505014}{{arXiv:nucl-ex/0505014}}}
\end{barticle}
\endbibitem

\bibitem[\protect\citeauthoryear{Lisa et~al.}{2000}]{Lisa:2000ip}
\begin{barticle}
\bauthor{\bsnm{Lisa}, \binits{M.A.}},
\bauthor{\bsnm{Heinz}, \binits{U.W.}},
\bauthor{\bsnm{Wiedemann}, \binits{U.A.}}:
\batitle{{Tilted pion sources from azimuthally sensitive HBT interferometry}}.
\bjtitle{Phys. Lett. B}
\bvolume{489},
\bfpage{287}--\blpage{292}
(\byear{2000})
\doiurl{10.1016/S0370-2693(00)00952-7}
{\href{https://arxiv.org/abs/nucl-th/0003022}{{arXiv:nucl-th/0003022}}}
\end{barticle}
\endbibitem

\bibitem[\protect\citeauthoryear{Mount et~al.}{2011}]{Mount:2010ey}
\begin{barticle}
\bauthor{\bsnm{Mount}, \binits{E.}},
\bauthor{\bsnm{Graef}, \binits{G.}},
\bauthor{\bsnm{Mitrovski}, \binits{M.}},
\bauthor{\bsnm{Bleicher}, \binits{M.}},
\bauthor{\bsnm{Lisa}, \binits{M.A.}}:
\batitle{{Correspondence between HBT radii and the emission zone in non-central
  heavy ion collisions}}.
\bjtitle{Phys. Rev. C}
\bvolume{84},
\bfpage{014908}
(\byear{2011})
\doiurl{10.1103/PhysRevC.84.014908}
{\href{https://arxiv.org/abs/1012.5941}{{arXiv:1012.5941}}}
{[nucl-th]}
\end{barticle}
\endbibitem

\bibitem[\protect\citeauthoryear{Lisa et~al.}{2011}]{Lisa:2011na}
\begin{barticle}
\bauthor{\bsnm{Lisa}, \binits{M.A.}},
\bauthor{\bsnm{Frodermann}, \binits{E.}},
\bauthor{\bsnm{Graef}, \binits{G.}},
\bauthor{\bsnm{Mitrovski}, \binits{M.}},
\bauthor{\bsnm{Mount}, \binits{E.}},
\bauthor{\bsnm{Petersen}, \binits{H.}},
\bauthor{\bsnm{Bleicher}, \binits{M.}}:
\batitle{{Shape analysis of strongly-interacting systems: The Heavy ion case}}.
\bjtitle{New J. Phys.}
\bvolume{13},
\bfpage{065006}
(\byear{2011})
\doiurl{10.1088/1367-2630/13/6/065006}
{\href{https://arxiv.org/abs/1104.5267}{{arXiv:1104.5267}}}
{[nucl-th]}
\end{barticle}
\endbibitem

\bibitem[\protect\citeauthoryear{Khyzhniak and Lisa}{2025}]{Khyzhniak:2024chj}
\begin{barticle}
\bauthor{\bsnm{Khyzhniak}, \binits{Y.}},
\bauthor{\bsnm{Lisa}, \binits{M.A.}}:
\batitle{{Pair momentum dependence of a tilted source in heavy-ion
  collisions}}.
\bjtitle{Phys. Rev. C}
\bvolume{111}(\bissue{2}),
\bfpage{024902}
(\byear{2025})
\doiurl{10.1103/PhysRevC.111.024902}
{\href{https://arxiv.org/abs/2410.15134}{{arXiv:2410.15134}}}
{[nucl-th]}
\end{barticle}
\endbibitem

\bibitem[\protect\citeauthoryear{Lisa et~al.}{2000}]{E895:2000opr}
\begin{barticle}
\bauthor{\bsnm{Lisa}, \binits{M.A.}}, \betal:
\batitle{{Azimuthal dependence of pion interferometry at the AGS}}.
\bjtitle{Phys. Lett. B}
\bvolume{496},
\bfpage{1}--\blpage{8}
(\byear{2000})
\doiurl{10.1016/S0370-2693(00)01280-6}
{\href{https://arxiv.org/abs/nucl-ex/0007022}{{arXiv:nucl-ex/0007022}}}
\end{barticle}
\endbibitem

\bibitem[\protect\citeauthoryear{Adamczewski-Musch
  et~al.}{2020}]{HADES:2019lek}
\begin{barticle}
\bauthor{\bsnm{Adamczewski-Musch}, \binits{J.}}, \betal:
\batitle{{Identical pion intensity interferometry at
  $\sqrt{s_{\mathrm{NN}}}=2.4~\hbox {GeV}$: HADES collaboration}}.
\bjtitle{Eur. Phys. J. A}
\bvolume{56}(\bissue{5}),
\bfpage{140}
(\byear{2020})
\doiurl{10.1140/epja/s10050-020-00116-w}
{\href{https://arxiv.org/abs/1910.07885}{{arXiv:1910.07885}}}
{[nucl-ex]}
\end{barticle}
\endbibitem

\bibitem[\protect\citeauthoryear{Abdallah et~al.}{2021}]{STAR:2021beb}
\begin{barticle}
\bauthor{\bsnm{Abdallah}, \binits{M.S.}}, \betal:
\batitle{{Global $\Lambda$-hyperon polarization in Au+Au collisions at $\sqrt
  {s_{NN}}$=3~GeV}}.
\bjtitle{Phys. Rev. C}
\bvolume{104}(\bissue{6}),
\bfpage{061901}
(\byear{2021})
\doiurl{10.1103/PhysRevC.104.L061901}
{\href{https://arxiv.org/abs/2108.00044}{{arXiv:2108.00044}}}
{[nucl-ex]}
\end{barticle}
\endbibitem

\bibitem[\protect\citeauthoryear{al.}{2026}]{STAR:2026skb}
\begin{barticle}
\bauthor{\bsnm{al.}, \binits{B.E.A.e.}}:
\batitle{Tilted geometry of the pion emission source in au+au collisions in the
  beam energy scan at the bnl relativistic heavy ion collider}.
\bjtitle{Phys. Rev. C}
(\byear{2026})
\doiurl{10.1103/zzcf-hysy}
{\href{https://arxiv.org/abs/2605.15013}{{arXiv:2605.15013}}}
{[nucl-ex]}
\end{barticle}
\endbibitem

\bibitem[\protect\citeauthoryear{Bass et~al.}{1998}]{Bass:1998ca}
\begin{barticle}
\bauthor{\bsnm{Bass}, \binits{S.A.}}, \betal:
\batitle{{Microscopic models for ultrarelativistic heavy ion collisions}}.
\bjtitle{Prog. Part. Nucl. Phys.}
\bvolume{41},
\bfpage{255}--\blpage{369}
(\byear{1998})
\doiurl{10.1016/S0146-6410(98)00058-1}
{\href{https://arxiv.org/abs/nucl-th/9803035}{{arXiv:nucl-th/9803035}}}
\end{barticle}
\endbibitem

\bibitem[\protect\citeauthoryear{Bleicher et~al.}{1999}]{Bleicher:1999xi}
\begin{barticle}
\bauthor{\bsnm{Bleicher}, \binits{M.}}, \betal:
\batitle{{Relativistic hadron hadron collisions in the ultrarelativistic
  quantum molecular dynamics model}}.
\bjtitle{J. Phys. G}
\bvolume{25},
\bfpage{1859}--\blpage{1896}
(\byear{1999})
\doiurl{10.1088/0954-3899/25/9/308}
{\href{https://arxiv.org/abs/hep-ph/9909407}{{arXiv:hep-ph/9909407}}}
\end{barticle}
\endbibitem

\bibitem[\protect\citeauthoryear{Becattini et~al.}{2013}]{Becattini:2013fla}
\begin{barticle}
\bauthor{\bsnm{Becattini}, \binits{F.}},
\bauthor{\bsnm{Chandra}, \binits{V.}},
\bauthor{\bsnm{Del~Zanna}, \binits{L.}},
\bauthor{\bsnm{Grossi}, \binits{E.}}:
\batitle{{Relativistic distribution function for particles with spin at local
  thermodynamical equilibrium}}.
\bjtitle{Annals Phys.}
\bvolume{338},
\bfpage{32}--\blpage{49}
(\byear{2013})
\doiurl{10.1016/j.aop.2013.07.004}
\end{barticle}
\endbibitem

\bibitem[\protect\citeauthoryear{Liu et~al.}{2026}]{Liu:2026lny}
\begin{botherref}
\oauthor{\bsnm{Liu}, \binits{D.-N.}},
\oauthor{\bsnm{Steinheimer}, \binits{J.}},
\oauthor{\bsnm{Sun}, \binits{K.-J.}},
\oauthor{\bsnm{Chen}, \binits{J.-H.}},
\oauthor{\bsnm{Ma}, \binits{Y.-G.}},
\oauthor{\bsnm{Bleicher}, \binits{M.}}:
{Hadron polarization and equation of state at FAIR/RHIC-BES energies}
(2026)
{\href{https://arxiv.org/abs/2606.13333}{{arXiv:2606.13333}}}
{[nucl-th]}
\end{botherref}
\endbibitem

\bibitem[\protect\citeauthoryear{Pratt et~al.}{1994}]{Pratt:1994uf}
\begin{barticle}
\bauthor{\bsnm{Pratt}, \binits{S.}}, \betal:
\batitle{{Testing transport theories with correlation measurements}}.
\bjtitle{Nucl. Phys. A}
\bvolume{566},
\bfpage{103}--\blpage{114}
(\byear{1994})
\doiurl{10.1016/0375-9474(94)90614-9}
\end{barticle}
\endbibitem

\bibitem[\protect\citeauthoryear{Aboona et~al.}{2026}]{STAR:2025njp}
\begin{barticle}
\bauthor{\bsnm{Aboona}, \binits{B.E.}}, \betal:
\batitle{{Measuring spin correlation between quarks during QCD confinement}}.
\bjtitle{Nature}
\bvolume{650}(\bissue{8100}),
\bfpage{65}--\blpage{71}
(\byear{2026})
\doiurl{10.1038/s41586-025-09920-0}
{\href{https://arxiv.org/abs/2506.05499}{{arXiv:2506.05499}}}
{[hep-ex]}
\end{barticle}
\endbibitem

\bibitem[\protect\citeauthoryear{Abdallah et~al.}{2023}]{STAR:2022fan}
\begin{barticle}
\bauthor{\bsnm{Abdallah}, \binits{M.S.}}, \betal:
\batitle{{Pattern of global spin alignment of {\ensuremath{\phi}} and K$^{*0}$
  mesons in heavy-ion collisions}}.
\bjtitle{Nature}
\bvolume{614}(\bissue{7947}),
\bfpage{244}--\blpage{248}
(\byear{2023})
\doiurl{10.1038/s41586-022-05557-5}
{\href{https://arxiv.org/abs/2204.02302}{{arXiv:2204.02302}}}
{[hep-ph]}
\end{barticle}
\endbibitem

\bibitem[\protect\citeauthoryear{Zhang et~al.}{2026}]{Zhang:2026tqs}
\begin{botherref}
\oauthor{\bsnm{Zhang}, \binits{K.}},
\oauthor{\bsnm{Wang}, \binits{X.}},
\oauthor{\bsnm{Luo}, \binits{X.}}:
{Spin Femtoscopy: A Framework for Revealing Genuine Spin Correlations}
(2026)
{\href{https://arxiv.org/abs/2607.00413}{{arXiv:2607.00413}}}
{[nucl-th]}
\end{botherref}
\endbibitem

\bibitem[\protect\citeauthoryear{Kiefer et~al.}{2026}]{Kiefer:2025xdp}
\begin{barticle}
\bauthor{\bsnm{Kiefer}, \binits{L.}},
\bauthor{\bsnm{Dash}, \binits{A.}},
\bauthor{\bsnm{Rischke}, \binits{D.H.}}:
\batitle{{Magnetization by rotation: Spin and chiral condensates in the NJL
  model}}.
\bjtitle{Phys. Rev. D}
\bvolume{113}(\bissue{9}),
\bfpage{094032}
(\byear{2026})
\doiurl{10.1103/t7wz-nvv9}
{\href{https://arxiv.org/abs/2509.18881}{{arXiv:2509.18881}}}
{[nucl-th]}
\end{barticle}
\endbibitem

\bibitem[\protect\citeauthoryear{Becattini et~al.}{2008}]{Becattini:2007sr}
\begin{barticle}
\bauthor{\bsnm{Becattini}, \binits{F.}},
\bauthor{\bsnm{Piccinini}, \binits{F.}},
\bauthor{\bsnm{Rizzo}, \binits{J.}}:
\batitle{{Angular momentum conservation in heavy ion collisions at very high
  energy}}.
\bjtitle{Phys. Rev. C}
\bvolume{77},
\bfpage{024906}
(\byear{2008})
\doiurl{10.1103/PhysRevC.77.024906}
\end{barticle}
\endbibitem

\bibitem[\protect\citeauthoryear{Becattini et~al.}{2013}]{Becattini:2013vja}
\begin{barticle}
\bauthor{\bsnm{Becattini}, \binits{F.}},
\bauthor{\bsnm{Csernai}, \binits{L.}},
\bauthor{\bsnm{Wang}, \binits{D.J.}}:
\batitle{{$\Lambda$ polarization in peripheral heavy ion collisions}}.
\bjtitle{Phys. Rev. C}
\bvolume{88}(\bissue{3}),
\bfpage{034905}
(\byear{2013})
\doiurl{10.1103/PhysRevC.88.034905} .
\bcomment{[Erratum: Phys.Rev.C 93, 069901 (2016)]}
\end{barticle}
\endbibitem

\bibitem[\protect\citeauthoryear{Becattini et~al.}{2017}]{Becattini:2016gvu}
\begin{barticle}
\bauthor{\bsnm{Becattini}, \binits{F.}},
\bauthor{\bsnm{Karpenko}, \binits{I.}},
\bauthor{\bsnm{Lisa}, \binits{M.}},
\bauthor{\bsnm{Upsal}, \binits{I.}},
\bauthor{\bsnm{Voloshin}, \binits{S.}}:
\batitle{{Global hyperon polarization at local thermodynamic equilibrium with
  vorticity, magnetic field and feed-down}}.
\bjtitle{Phys. Rev. C}
\bvolume{95}(\bissue{5}),
\bfpage{054902}
(\byear{2017})
\doiurl{10.1103/PhysRevC.95.054902}
\end{barticle}
\endbibitem

\bibitem[\protect\citeauthoryear{He et~al.}{2020}]{HeJJ:2020}
\begin{barticle}
\bauthor{\bsnm{He}, \binits{J.J.}},
\bauthor{\bsnm{Zhang}, \binits{S.}},
\bauthor{\bsnm{Ma}, \binits{Y.G.}},
\bauthor{\bsnm{Chen}, \binits{J.H.}},
\bauthor{\bsnm{Zhong}, \binits{C.}}:
\batitle{{Clustering structure effect on Hanbury-Brown{\textendash}Twiss
  correlation in $^{12}\hbox {C} {+^{197}}\hbox {Au}$ collisions at 200 GeV}}.
\bjtitle{Eur. Phys. J. A}
\bvolume{56},
\bfpage{52}
(\byear{2020})
\doiurl{10.1140/epja/s10050-019-00002-0}
\end{barticle}
\endbibitem

\bibitem[\protect\citeauthoryear{Wang et~al.}{2025}]{WangDF:2025}
\begin{barticle}
\bauthor{\bsnm{Wang}, \binits{D.-F.}},
\bauthor{\bsnm{Chen}, \binits{M.-Y.}},
\bauthor{\bsnm{Ma}, \binits{Y.-G.}},
\bauthor{\bsnm{Shou}, \binits{Q.-Y.}},
\bauthor{\bsnm{Zhang}, \binits{S.}},
\bauthor{\bsnm{Zheng}, \binits{L.}}:
\batitle{Investigating the pion emission source in pp collisions using the ampt
  model with subnucleon structure}.
\bjtitle{Nucl. Sci. Tech.}
\bvolume{36},
\bfpage{154}
(\byear{2025})
\doiurl{10.1007/s41365-025-01722-3}
\end{barticle}
\endbibitem

\end{thebibliography}

\appendix
\newpage
\clearpage

\noindent
{\large \textbf{Methods}}
\vspace{6pt}

\bmhead{Global $\Lambda$ polarization}\label{methods:polarization}
We employ the Ultra-relativistic Quantum Molecular Dynamics (UrQMD) transport model~\cite{Bass:1998ca,Bleicher:1999xi}, with an equation of state from hadron resonance gas (HRG) model, to describe Au+Au collisions at $\sqrt{s_{\rm NN}} = 2.24$–$7.7\text{ GeV}$. The UrQMD model   has been extensively applied to the study of particle production and collective dynamics in these  collision energies.

The $\Lambda$ polarization can be determined from the thermal vorticity under the assumption of local thermal equilibrium~\cite{Becattini:2007sr,Becattini:2013vja,Becattini:2016gvu}. 
To the first order in thermal gradients, the mean polarization four-vector of a spin-1/2 $\Lambda$ hyperon emitted at space-time point $x$ with four-momentum $p$ is given by~\cite{Becattini:2013fla} 
\begin{equation}
\mathcal{P}^\mu(x,p) = 
-\frac{1}{4m}(1-n_F)\epsilon^{\mu\nu\rho\sigma}p_\nu\varpi_{\rho\sigma}(x)
\label{eq:pol_vec}
\end{equation}
where $m$ is the $\Lambda$ mass and $n_F$ is the Fermi--Dirac distribution with $1-n_F\approx 1$ \cite{Becattini:2013vja}. In Eq.~\eqref{eq:pol_vec}, the thermal vorticity tensor is defined as $
\varpi_{\mu\nu}=-\left(\partial_\mu \beta_\nu-\partial_\nu \beta_\mu\right)/2$
where $\beta_\mu \equiv u_\mu/T$  with $T$ being temperature and $u^\mu$ being the fluid four velocity satisfying $u_\mu u^\mu=1$. For approximately constant temperature,   $\varpi_{\mu\nu}\simeq\omega_{\mu\nu}/T$ where
$\omega_{\mu\nu}=-(\partial_\mu u_\nu-\partial_\nu u_\mu)/2$
is  the kinetic vorticity tensor.

We evaluate the thermal and kinetic vorticities from the UrQMD particle phase-space distributions at the kinetic freeze-out where particles experience their last collisions. Events are averaged and coarse-grained on a four-dimensional spacetime lattice with spatial spacing $\Delta x=\Delta y=\Delta z=1~\mathrm{fm}$ over $(-50,50)~\mathrm{fm}$ and temporal spacing $\Delta t=1~\mathrm{fm}/c$ up to $50~\mathrm{fm}/c$. In each cell, the local energy density, net-baryon density, and momentum density are calculated to determine the temperature $T$ and flow four-velocity $u^\mu$ through the HRG equation of state~\cite{Liu:2026lny}. The global $\Lambda$ polarization is evaluated in the particle's rest frame with a Lorentz boost~\cite{Becattini:2013fla}, where $\mathcal{P}^\mu=(0, \vect{\mathcal{P}}^*)$. It is then obtained by projecting the spin polarization three-vector $\vect{\mathcal{P}}^*$ onto the direction of the orbital angular momentum  $\vect J_\text{sys}$, i.e.,
\begin{equation}
    \mathcal{P}_\Lambda=\langle \frac{ \vect {\mathcal{P}}^*\cdot \vect J_\text{sys}  }{|\vect J_\text{sys}|}\rangle
\end{equation}
where the angular bracket denotes the ensemble average over events in certain centrality class.

\vspace{6pt}

\bmhead{Source tilt angle}\label{methods:tilt}
The source geometry can be accessed through azimuthally sensitive HBT interferometry with identical pion pairs~\cite{Lisa:2000ip,Lisa:2005dd, HeJJ:2020,WangDF:2025}. At mid-rapidity, the azimuthal angle dependence of the extracted squared HBT radii from Eq. (\ref{eq:hbtcorr}) can be decomposed into Fourier harmonics~\cite{Lisa:2000ip},
\begin{equation}
\begin{aligned}
R_{{o }}^{2}(\phi)
&= R_{{o },0}^{2}
+ 2 \sum_{n=2,4,6,\ldots}
R_{{o },n}^{2}\cos(n\phi), \\
R_{{s }}^{2}(\phi)
&= R_{{s },0}^{2}
+ 2 \sum_{n=2,4,6,\ldots}
R_{{s },n}^{2}\cos(n\phi), \\
R_{{l }}^{2}(\phi)
&= R_{{l },0}^{2}
+ 2 \sum_{n=2,4,6,\ldots}
R_{{l },n}^{2}\cos(n\phi), \\
R_{{os}}^{2}(\phi)
&= 2 \sum_{n=2,4,6,\ldots}
R_{{os},n}^{2}\sin(n\phi), \\
R_{{ol}}^{2}(\phi)
&= 2 \sum_{n=1,3,5,\ldots}
R_{{ol},n}^{2}\cos(n\phi), \\
R_{{sl}}^{2}(\phi)
&= 2 \sum_{n=1,3,5,\ldots}
R_{{sl},n}^{2}\sin(n\phi).
\end{aligned}
\label{eq:hbt_radii}
\end{equation}
The source dimensions in the reaction plane can be determined from the Fourier coefficients in Eq.~\eqref{eq:hbt_radii}, i.e,
\begin{equation}
    \begin{aligned}
        \langle\tilde x_i^2\rangle&=R^2_{s,0}-2R^2_{s,2},\\
         \langle\tilde z_i^2\rangle&=R^2_{l,0},\\
         \langle\tilde x_i \tilde z_i\rangle&=-2R^2_{sl,1}.
    \end{aligned}
    \label{eq:}
\end{equation}
where $\tilde x=x-\avg{x}$ and $\tilde z=z-\avg{z}$ with angular brackets denoting the average over the freeze-out hypersurface, and the subscripts denote the corresponding Fourier harmonics. The tilt angle $\theta_s$ is then obtained from the covariance matrix of the emission points in the $z$-$x$ plane~\cite{Lisa:2000ip,Lisa:2011na},
\begin{equation}
\begin{aligned}
    \theta_s
&=\frac12\tan^{-1}\left(\frac{2\langle\tilde{x}\tilde{z}\rangle}{\langle\tilde{z}^2\rangle-\langle\tilde{x}^2\rangle}\right),\\
    &\approx\frac{1}{2}\tan^{-1}\!\left(\frac{-4 R_{sl,1}^2}{R_{l,0}^2-R_{s,0}^2+2  R_{s,2}^2}\right),
\end{aligned}
\label{eq:tilt_angle}
\end{equation}
a relation that is widely used in experimental analyses.

The relation in Eq. (\ref{eq:tilt_angle}) can be verified for a Gaussian ellipsoid source that is parametrized in space and time as 
$S\propto\exp[ 
        -\frac{(x\cos\theta_s-z\sin\theta_s)^2}{2\sigma_{x'}^2}  -\frac{y^2}{2\sigma_{y'}^2}
         -\frac{(x\sin\theta_s+z\cos\theta_s)^2}{2\sigma_{z'}^2}
        -\frac{t^2}{2\sigma_t^2} ]$
where $\sigma_t$ denotes the emission duration, $\sigma_{x’,y',z'}$ denotes  the width of the tilted ellipsoid with primes corresponding to principal axes, and the spatial tilt angle $\theta_s$ characterizes the angle between the  beam direction and the major axis of the emission ellipsoid in the reaction plane.

\vspace{6pt}
\begin{figure}[!t]
    \centering
    \begin{subfigure}[t]{0.53\columnwidth}
        \centering
        \includegraphics[width=\linewidth]{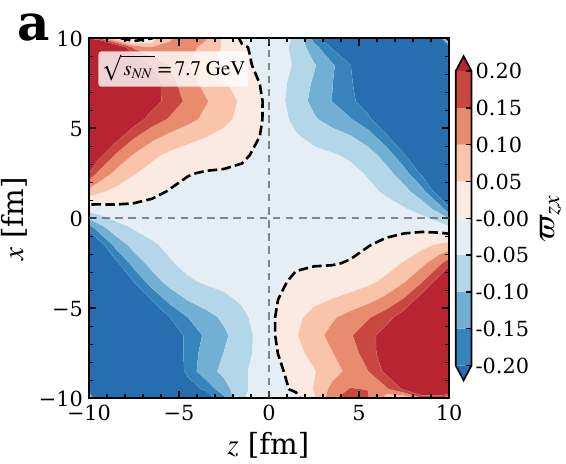}
    \end{subfigure}
    \hfill
    \begin{subfigure}[t]{0.42\columnwidth}
        \centering
        \includegraphics[width=\linewidth]{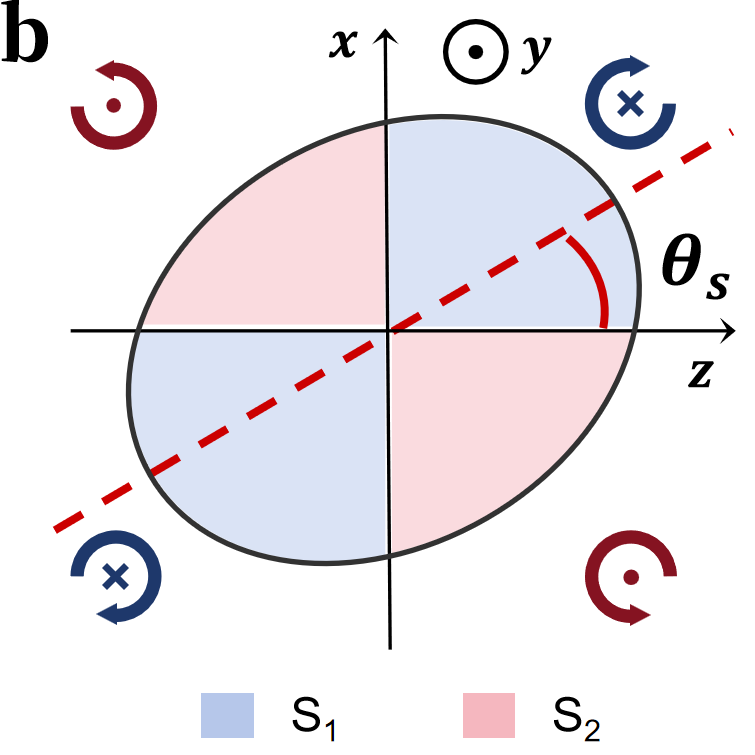}
    \end{subfigure}

    \caption{\textbf{Thermal vorticity and its quadrupolar structure in the reaction plane.}
    \textbf{(a)} Distribution of the thermal vorticity component $\varpi_{zx}$ 
    in the reaction plane from UrQMD simulations of Au+Au collisions at 
    $\sqrt{s_{\rm NN}}=7.7$ GeV in the 30--40\% centrality class at 
    $t=15$ fm/$c$. 
    \textbf{(b)} Schematic of the tilted source and quadrupolar vorticity 
    in the reaction plane. The fireball is depicted as an ellipse tilted by 
    an angle $\theta_s$, with the vorticity exhibiting a quadrupolar pattern.}
    \label{fig:vorticity}
\end{figure}

\bmhead{Relation between $\mathcal{P}_\Lambda$ and $\theta_s$}\label{methods:linear}
To give an intuitive understanding about the connection between global $\Lambda$ polarization and source tilt angle, we display in panel (a) of Fig.~\ref{fig:vorticity}  the thermal vorticity distribution  in the $z$--$x$ plane at 15~fm/$c$ for Au+Au collisions at $\sqrt{s_{NN}}=7.7$~GeV.  The vorticity exhibits a quadrupole pattern in the reaction plane, whose sign alternates across the four quadrants, as also found in Refs.~\cite{Li:2017slc,Shi:2017wpk}. This pattern arises because the transverse velocity $v_x$ decreases with increasing distance from $z=0$ along the longitudinal direction.
This non-uniform field motivates the  
following illustrative source-weighting argument. Consider a centered elliptical source in the reaction plane and the magnitude of the thermal vorticity, denoted by $|\varpi_{zx}|$, is the same in each quadrant (Fig.~\ref{fig:vorticity} (b)).    Let $S_1$ be the combined source area in the first and third quadrants and $S_2$ that in the second and fourth quadrants. In this approximation, the $\Lambda$ polarization is proportional to the area difference of $S_1-S_2= 2(R_a^2-R_b^2)\theta_s$ where $R_a$ and $R_b$ are the semi-major and semi-minor axes of the source, respectively.
Equation~\eqref{eq:pol_vec} then relates the global polarization to the source-averaged vorticity by
\begin{equation}
    \mathcal{P}_\Lambda = \frac{E}{2m}\frac{2(R_a^2-R_b^2)}{\pi R_aR_b}|\varpi_{zx}|\theta_s,
    \label{eq:pola_tilt2}
\end{equation}
where  $E$ and $m$ are the $\Lambda$ energy and mass, respectively.  The global $\Lambda$ polarization  thus depends on  both the local thermal vorticity  and the source geometry. Using representative parameters, $E/m\approx1.7$, $R_a/R_b\approx2.2$, and $|\varpi_{zx}|\approx0.07$ for semi-central collisions, Eq.~\eqref{eq:pola_tilt2} yields 
$\mathcal{P}_\Lambda\approx0.066\theta_s$ which is consistent with theoretical calculations by UrQMD and experimental measurements shown in Fig.~\ref{fig:barnett}. 

Equation~\eqref{eq:pola_tilt2} provides an intuitive small-angle estimate, which neglects spatial and temporal variations of the vorticity, realistic emission weights, and the full momentum dependence of the rest-frame polarization. In the numerical UrQMD calculations, the polarization is instead evaluated directly from the local fields at the hyperon emission points and boosted to the hyperon rest frame. Thus, the analytical relation provides physical insight into the polarization--geometry connection, while the transport calculation establishes it quantitatively across the studied energy range.


\end{document}